\documentstyle[aps,preprint,epsfig]{revtex}
 
\def\be{\begin{eqnarray}}
\def\ee{\end{eqnarray}}
\def\nn{\nonumber}
\def\eq#1{(\ref{#1})}

\def\({\left(}
\def\[{\left[}
\def\){\right)}
\def\]{\right]}
\def\h{{1\over 2}}
\def\.{\cdot}
\def\labels#1{\label{#1}}

\def\s{\sigma}
\def\e{\epsilon}
\def\d{\delta}

\begin{document}

\title{SMALL-RECOIL APPROXIMATION}
\author{C.S. Lam}
\address{Department of Physics, 
McGill University\\
3600 University St., Montreal, QC,
Canada H3A 2T8\\
email: Lam@physics.mcgill.ca}

\maketitle

\begin{abstract}
In this review we discuss a technique to compute
and to sum a class of Feynman diagrams, and some of its 
applications. These are diagrams
containing one or more energetic particles that suffer very
little recoil in their interactions. When recoil is completely neglected,
a decomposition formula can be proven.
This formula is a generalization of the well-known eikonal
formula, to non-abelian interactions. It expresses the amplitude
as a sum of 
products of the irreducible amplitudes, with each irreducible
amplitude being the amplitude to emit
one, or several mutually interacting, quasi-particles.
For abelian interaction a quasi-particle is nothing but
the original boson, so this decomposition formula reduces
to the eikonal formula. In non-abelian situations each
quasi-particle can be made up of many bosons, though always with
a total quantum number identical to that of a single boson. 
This decomposition enables certain amplitudes of all orders to be summed
up into an exponential form, and it allows subleading contributions
of a certain kind, which is difficult to reach in the usual
way, to be computed. For bosonic emissions from a heavy source with many 
constituents, a quasi-particle amplitude turns out to be
an amplitude in which all bosons are emitted from the same constituent.
For high-energy parton-parton scattering in the near-forward
direction, the quasi-particle turns out to be the Reggeon, and this
formalism shows clearly why gluons reggeize but photons do 
not. The ablility to compute subleading terms in this formalism
 allows the BFKL-Pomeron amplitude to be extrapolated
to asymptotic energies, in a unitary way preserving the Froissart
bound. We also consider recoil corrections for abelian interactions
in order to accommodate the Landau-Pomeranchuk-Migdal effect.
\end{abstract}

\section{Introduction}
Emission and absorption of soft particles 
cause hardly any recoil to an energetic source.
Surprisingly, this trivial fact gives rise to
a great deal of simplification in quantum field theoretical calculations. 
This is the subject we wish to review in the article.

Such a source with an energy $p^0=\sqrt{\vec p^2+m^2}$ may be 
{\it relativistic} when $|\vec p|$ is large and $m$ is small, or
{\it non-relativistic} when $|\vec p|$ is small and $m$ is large.
For example, a high-energy quark is a relativistic source of gluons,
and a heavy nucleus at rest is a non-relativistic source of soft pions.

The origin of this simplification is a {\it decomposition formula} 
\cite{LL1LJMP} for the
tree amplitude $A_n$ of $n$ identical bosons. This formula allows 
$A_n$ to be decomposed into a sum of products of {\it irreducible 
amplitudes} $I_m$, where $m$ runs from 1 to $n$, labeling the
number of bosons in it.  The number $k$
of irreducible factors in each term again varies from 1 to $n$, 
but to preserve the total number of bosons emitted the condition 
$\sum_{i=1}^k m_i=n$ must be obeyed. An irreducible amplitude $I_m$
differs from a Feynman tree amplitude $A_m$ only in having the product
of vertex factors $V_1V_2\cdots V_{m-1}V_m$ in $A_m$ replaced by their
nested commutator $[V_1,[V_2,[\cdots,[V_{m-1},V_m]\cdots]]]$ in $I_m$.
The decomposition formula is combinatorial in nature
and will be
discussed more fully in Sec.~2. It is valid whatever
$V_i$'s are, and whether the bosons are on-shell or off-shell.
The latter makes it possible for the tree diagram 
in question to be a part of 
a much larger Feynman diagram, thus allowing
 the decomposition of tree amplitudes
to be applied fruitfully to loop diagrams as well.

The vertex factors $V_i$ are matrices causing a change in the
spin and internal quantum numbers of the energetic source
after each emission (or absorption). They are {\it abelian} if
the spin and quantum numbers remain the same after each emission. In
that case they are  diagonal and commute with one another, 
so
all $I_m$ vanish except $I_1=A_1$. As a result, the 
$n$-boson amplitude $A_n$ is factorized into a product of $n$
single-boson amplitudes $A_1$. Such a decomposition for abelian vertices
have been known for a long time  under the name
of an {\it  eikonal formula} \cite{EIK}. It is often used to demonstrate
the cancelation of infrared divergence in 
QED \cite{STERMAN,PESKIN,WEINBERG}, and to establish the geometrical nature
of a scattering amplitude at high energy \cite{GLAUBER,CW}. It is well 
documented in text books so we shall not discuss it any further
until Sec.~5, when its recoil correction is considered.

Let us examine more carefully what decomposition means 
in the non-abelian context.
We shall concentrate on the predominant
situation when the vertices $V_i$ are generators of a Lie group. 
In that case
the emitted bosons carry the quantum numbers of the adjoint representation.
Since a nested commutator of generators is a generator, the $m$
bosons emitted in $I_m$ also carry a total  quantum number 
in the adjoint representation.
We may therefore think of these $m$ bosons together
to form a {\it quasi-particle},
with the same quantum number as an original boson.
The irreducible amplitude $I_m$ is then an amplitude for 
the emission of a single
quasi-particle. For that reason we shall use the terms `irreducible
amplitude' and `quasi-particle amplitude' interchangeably.
The decomposition formula then says 
multi-quasi-particle amplitudes are always factorizable into 
products of single-quasi-particle amplitudes, much like the abelian
case. Thus the decomposition formula is also referred to as 
the {\it factorization
formula}. In this language the only difference between abelian and
non-abelian vertices is that a quasi-particle in the abelian case is
the original boson, whereas in the non-abelian case it may have a 
complicated structure consisting of many bosons. This 
difference is eventually responsible for the gluon
to reggeize and not the photon, because the quasi-particle
in high-energy scattering turns out to be just the Reggeon. For more
details see Sec.~4.

The decomposition formula is able to simplify
calculation in at least three ways.
First, all the terms in $A_n$ with $k>1$ can be computed from
$I_m$ with $m<n$, which in turn can be computed from $A_m$ with $m<n$. 
The only new term needed to be computed
at order $n$ is the quasi-particle amplitude $I_n$, thus
reducing the labor of computation.
Secondly, factorization often allows $A_n$ of all $n$ to be summed
up, typically into an exponential function of $I_m$. This is important
in problems where phase is of paramount concern, as  
in the unitarization of total cross section in
Sec.~4, and the coherent reduction of radiation in the
Landau-Pomeranchuk-Migdal (LPM) effect to be discussed later 
and in Sec.~5.
Thirdly, 
quite often all $I_m$ have the same order of magnitude $\mu$, 
irrespective of what $m$ is. This will be the case for the examples
discussed in Secs.~3 and 4.
In that case
terms with the product of $k$ irreducible amplitudes
are of order $\mu^k$, and it is important to note also that
these are also terms with quantum numbers
made up of $k$ adjoint objects. 
This correlation between the magnitude
and the quantum number is useful as we shall illustrate in Sec.~3.
In particular, when $\mu\ll 1$, this means that 
magnitude is determined by 
the quantum number. To see that let ${\cal C}_k$ be the set of 
internal quantum numbers first
appearing in the tensor product of $k$ adjoint objects.
For example, in the case when 
the internal quantum number is $SU(3)$ color,
octet alone is in ${\cal C}_1$, whereas singlets,
deculplets, anti-decuplets, and 27-plets are in ${\cal C}_2$.
In this notation,
amplitudes belonging to ${\cal C}_k$ is of order $\mu^k$
when $\mu\ll 1$. Though small, such subdominant amplitudes can be
extracted in this approach
simply by multiplying $k$ dominant amplitudes
$I_m$ together, each of order $\mu$. This is not something that
can be obtained in the usual way 
short of carrying out  the 
very difficult task of calculating each Feynman
diagram to an accuracy of $\mu^k$. These subdominant contributions
are needed in Sec.~4 to preserve the Froissart bound when
total cross section is extrapolated 
to asymptotic energies.

These advantages of the decomposition formula will be
illustrated with two concrete examples. In Sec.~3, we consider a
source made up of $N\gg 1$ constituents. That source
 may be a
heavy nucleus emitting soft gluons \cite{MV,LM}, or a baryon emitting
soft pions in a QCD theory with $N\gg 1$ colors \cite{LARGEN}. The magnitude of
each of the $n!$ Feynman diagram making up $A_n$ is of order
$N^n$, since each boson can be emitted from any of the $N$ constituents.
It will be shown that the irreducible amplitude $I_n$ corresponds to
emission of all $n$ pions from the {\it same} constituent, hence
$I_n$ is of order $N\equiv\mu$, independent of $n$. 
Unless there are selection rules forbidding it, the individual bosons
would like to be emitted from as many different constituents as possible
to maximize the matrix element. This leads to the classical approximation
used in the McLerran-Venugopalan model \cite{MV}. 
In the case when $A_n$ 
describes a scattering amplitude with one incoming and $n-1$ outgoing
pions, terms in the decomposition with $k>1$ vanish. This is so
because at least 
one of the $k$ irreducible factors must contain only outgoing pions,
and by energy conservation this cannot survive. Since the only
surviving term has $k=1$, $A_n$ is of order 
$\mu=N$ \cite{LL2},
rather than the magnitude $N^n$ of each of its Feynman diagrams.
This enormous cancelation even for a moderate $n$ is not
something one can easily obtained by calculating 
directly from the Feynman diagrams.

In Sec.~4 we consider parton diffractive scattering in perturbative
QCD, in the limit of small coupling $\alpha_s\ll 1$, and high energy
$\sqrt{s}$ so that $\alpha_s\ln s=O(1)$. The scattering
amplitude $\widetilde A(s,b)$ (where $b$ is the impact
parameter) is no longer a tree, but
it can still be decomposed into irreducible amplitudes $\delta_k(s,b)$
using the tree decomposition for each of the two fast partons. 
The irreducible amplitude $\delta_k$ is 
characterized by the exchange
of $k$ mutually interacting
quasi-particles, and is of order $\alpha_s^k$, so 
once again we have a factor $\mu=\alpha_s$ per quasi-particle.
This observation will be used to devise a way to unitarize the
BFKL-Pomeron amplitude \cite{BFKL}, 
so that the total cross section when extrapolated to asymptotic
energies obey the Froissart  bound \cite{DKL}.

So far we have neglected the recoil of the energetic source.
We will consider how to include it in the abelian situation
 in Sec.~5. This is prompted by the 
Landau-Pomeranchuk-Migdal (LPM) effect in
QED \cite{LPM}, which describes a suppression of radiation for a particle
traversing in a dense medium. More specifically, consider an 
energetic electron 
with energy $E=p_1^0$ moving in
a medium of scatterers spaced apart by a distance $d$.
Let the energy of the emitted photon be $\omega=k^0=xE$, 
and the energy of the final electron be
$p_2^0=(1-x)E$. 
Neglecting the mass of the charged particle, the longitudinal
momentum transfer in such a process is $q^3=p_2^3+k^3-p_1^3
\simeq(p_{2\perp}^2/(1-x)+k_{\perp}^2/x)/(2E)$. If $|q^3d|\ll 1$,
emission from different scatters is coherent, as if it came from 
one source instead of  many. 
This causes a suppression of radiation
compared to the case when the emission is incoherent, when
 the intensity
of emission is proportional to the number of scatters. This suppression
is the LPM effect. Since $q^3=O(1/E)$, we must also compute the 
influence of scatters to $O(1/E)$ for consistency, and as we shall see
in Sec.~5, this calls for a recoil correction to the same order.

\section{Recoilless Emission}
\subsection{Propagator}
Consider a relativistic source moving 
parallel to the z-axis at nearly the speed of light. 
Its transverse position ${x_\perp}$ and its lightcone distance
$x^-\equiv x^0-x^3$ are fixed provided the emission of particles
causes no recoil. This classical picture holds
even quantum mechanically for the following reason.
Let ${p'}^\alpha$ be the on-shell momentum of the energetic source
after all emissions ($p\.p=m^2$), and $k^\alpha_i$ the outgoing 
momentum of the $i$th boson (Fig.~1), with $|k^\alpha_i|\ll {p'}^0$ for every
$\alpha$ and every $i$. Then the propagators of Fig.~1 can be
approximated by
\be
P(p'+K)\equiv{1\over (p'+K)^2-m^2+i\e}\simeq {1\over 2p'\.K+i\e},
\labels{prop}\ee
where $K$ is an appropriate sum of $k_i$'s and the term
$K\.K$ has been neglected compared to $2p'\.K$. The propagator
in configuration space is obtained by Fourier transform to be
\be
\widetilde P(x)&\equiv&{1\over (2\pi)^4}\int d^4(p'+K)e^{-i(p'+K)\.x}
{1\over (p'+K)^2-m^2+i\e}\nn\\
&\simeq& -{i\over {p'}^+}e^{-ip'\.x}
\d^2({x_\perp})\d(x^-)\theta(x^+),\labels{propx}\ee
where the lightcone coordinates are defined by $A^\pm=A^0\pm A^3$.
With this definition and with the mass of the source neglected,  the 
only non-zero component of the source momentum may
be taken to be  ${p'}^+\equiv 2E$. When the source
particle propagates from $x_1$ to $x_2$ with $x=x_1-x_2$, this expression
confirms that the transverse and the lightcone positions
of the particle are fixed, while its lightcone time $x^+$ goes forward
as it propagates, as in the classical picture.

Note that if the Feynman propagator in \eq{prop} is replaced by the
Cutkosky propagator $-2\pi i\delta[(p+K)^2-m^2]$, then the factor
$\theta(x^+)$ in \eq{propx} is replaced by 1.

For a non-relativistic source with $|K^\mu|\ll m$, the corresponding
propagator becomes
\be
-{i\over 2m}e^{-imx^0}\d^3(\vec x)\theta(x^0).\labels{nrpropx}\ee

\subsection{Decomposition and Factorization}
To conform to Bose-Einstein statistics, the emitted bosons in Fig.~1
must be symmetrized, with the full $n$-boson 
amplitude $A_n$ given by a sum of $n!$ permuted diagrams.
Labeling these permuted diagrams by the order of their bosonic emissions,
as in $[\s]=[\s_1\s_2\cdots \s_n]$, so that Fig.~1 is $[12\cdots n]$,
we have
\be
A_n=\sum_{[\s]\in S_n}A[\s]=\sum_{[\s]\in S_n}V[\s]P[\s],
\labels{an}\ee
where the sum is taken over the set $S_n$ of all permutations
 of $n$ objects.
Each $A[\s]$ is given by the product of the vertex factor
$V[\s]=V_{\s_1}V_{\s_2}\cdots V_{\s_n}$, and the product $P[\s]$
of all the propagators. 
If the initial source particle is off-shell, 
there are $n$ propagators
like \eq{prop}.
If it is on-shell, then there are only $n-1$ propagators, but it is
convenient to include also an explicit on-shell $\delta$-function
factor $-2\pi i\delta(p'\.p'-m^2)$ into $A_n$, where $p'=p+\sum_{i=1}^nk_i$
is the momentum of the initial source particle. In the no-recoil
approximation, the argument of this Cutkosky propagator becomes
 $p'\.p'-m^2\simeq 2p\.\sum_{i=1}^nk_i$. 

In configuration space, each Feynman propagator is given by \eq{propx}.
The Cutkosky propagator is given by the same formula with $\theta(x^+)$
replaced by 1.  We shall denote
the product of propagators in the configuration space by 
$\widetilde P[\s]$, so that the configuration-space amplitude is
\be
\widetilde A_n=\sum_{[\s]\in S_n}V[\s]\widetilde P[\s].
\labels{anx}\ee
To establish the decomposition formula, the most important factor
in  $\widetilde P[\s]$ is
$\theta[\s]=\prod_{i=1}^{n}\theta(x_i^+-x_{i+1}^+)$,
where $x_{n+1}\equiv x_0$ is the spacetime coordinate of the initial
source if it is off-shell, and it is $-\infty$ if it is on-shell. 
We shall denote the amplitude by $\widetilde A'_n$ when $\widetilde P[\s]$
is replaced by $\theta[\s]$.

Before we proceed
note that the propagator in \eq{prop} contains no numerator
factor. This implies that either helicity is conserved
as in a vector-boson emission from a relativistic source, or that
the spin-flip information is encoded in the vertex factor $V_i$
as in non-relativistic Yukawa interaction. To illustrate how the
numerator can formally be made to disappear, suppose the energetic source to
be a spin-$\h$ particle.
If $p+K$ is the momentum of the propagator, as
in \eq{prop}, then the numerator is
$\gamma\.(p+K)+m\simeq \gamma\.p=\sum_\lambda
u_\lambda(p)\bar u_{\lambda}(p)$. We then move the $u$ factor leftward
and the $\bar u$ factor rightward, and absorb them into the vertex
factors $V_{\s_i}$. In this way the numerator factor can be made to
disappear.

To explain the decomposition formula, look at the simplest case when $n=2$:
\be
\widetilde A_2'&=&\theta[12]V_1V_2+\theta[21]V_2V_1\nn\\
&=&(\theta[12]+\theta[21])V_1V_2+\theta[21][V_2,V_1]\nn\\
&=&\(\theta[1]V_1\)
\(\theta[2]V_2\)+\theta[21][V_2,V_1],
\labels{b2}\ee
where the identity
$\theta[12]+\theta[21]=\theta[1]\theta[2]\equiv \theta(x_1^+-x_0^+)
\theta(x_2^+-x_0^+)$ has been used. We shall now write this formula
in a way that can be generalized to all $n$. 

For that purpose
introduce the {\it quasi-particle amplitude}
\be
\widetilde Q'[\s]&=&\theta[\s][V_{\s_1},[V_{\s_2},[\cdots,[V_{\s_{n-1}},V_{\s_n}]\cdots]]],
\labels{nc}\ee
which is obtained from the Feynman amplitude
$\widetilde A'[\s]$ by replacing  the products
in $V[\s]$ by their nested commutators. If $\s',\s'',\cdots,\s'''$
are sequences of non-overlapping numbers, we will also let
\be
\widetilde Q'[\s'|\s''|\cdots|\s''']=\widetilde Q'[\s']
\widetilde Q'[\s'']\cdots \widetilde Q'[\s'''].
\labels{fact}\ee
In other words, vertical bars (or {\it cuts}) are used to
separate the factors into products. 

With this notation, we can rewrite \eq{b2} to be
\be
\widetilde A_2'&=&\widetilde Q'[1|2]+\widetilde Q'[21].
\labels{a2}\ee
The general decomposition formula is given by \cite{LL1LJMP}
\be
\widetilde A_n'&=&\sum_{[\s]\in S_n}\widetilde Q_{c}'[\s],
\labels{decomp}\ee
where $\widetilde Q_c'[\s]$ is by definition
equal to $\widetilde Q'[\sigma]$
with vertical bars suitably inserted 
into the argument $[\sigma]$.
The rule for insertion is the following:
a vertical bar is inserted after the number $\s_i$ in $[\s]$
if and only if there is no number to its right smaller than it.
Thus for example $\widetilde Q'_c[12]=\widetilde Q'[1|2]$,
and $\widetilde Q'_c[21]=\widetilde Q'[21]$, so \eq{a2} is a special
case of \eq{decomp}.
More
complicated examples for $n=8$ are:
 $\widetilde Q'_c[64312857]=
\widetilde Q'[6431|2|85|7]= Q'[6431]Q'[2]Q'[85]Q'[7]$,
and $Q'_c[12385476]=Q'[1|2|3|854|76]=Q'[1]Q'[2]Q'[3]Q'[854]Q'[76]$.
These are illustrated graphically in Fig.~2, 
where a cut denotes factorization in the configuration space. 
We have also shown simplified versions where each quasi-particle is indicated
by a single thick vertical line.
Note however from the remark in the
paragraph following \eq{propx} that in momentum space,
a vertical means replacing
the Feynman propagator by the corresponding Cutkosky propagator.

To obtain the decomposition formula for 
the full configuration-space amplitude
$\widetilde A_n$, we simply remove the prime on both sides
of \eq{decomp}. To get the decomposition formula for the momentum-space
amplitude $A_n$, we will remove the tilde on both sides as well.
With or without tilde and/or prime, $Q_{c}$ 
 is always related to
$Q$ by the same rule of insertion of vertical bars. It is the
multi-quasi-particle amplitude (also known as the
{\it non-abelian cut amplitude}) that factorizes according to the
vertical bars into products of the irreducible single-quasi-particle
amplitudes $Q[\s]$.
The irreducible amplitude $Q[\s]$ (without vertical bars)
is always obtained from the Feynman amplitude
$A[\s]$ by replacing the products of $V_i$'s
with their nested commutators. These are the quantities $I_m$ mentioned
in the Introduction, with $m$ being the length of $[\s]$.

It is easy to incorporate the decomposition formula into
Feynman diagrams by a slight modification of the latter.
All that one has to do is to add cuts to the energetic tree(s) according
to the insertion rules. Figs.~2 and 4 are examples of such
{\it non-abelian cut diagrams} \cite{FHL}. 
Ordinary Feynman rules remain unchanged,
except for two modifications. A cut propagator is a Cutkosky propagator
instead of a Feynman propagator, and the product of vertex factors
$V_i$ between cuts are changed into their nested commutators.
Note that individual Feynman diagrams are not the same as individual
cut diagrams, but according to \eq{an} and \eq{decomp} their permuted
sums are equal.

\section{Emission from a Composite Source}
Consider a non-relativistic source 
made up of $N$ constituents, with 
$\psi^\dagger,\psi$ being the creation and annihilation operators for these
constituents. Bosons interact with the source through
a vertex of the form $V_i=\lambda\psi^\dagger\Gamma_i\psi$.
The matrix element of $V_i$ is of order $\lambda N$, 
since each boson can
be emitted from any of the $N$ constituents of the source. 
See Fig.~3. If we have a product
of $m$ such operators, $V_1V_2\cdots V_m$, then its matrix element is
of order $(\lambda N)^m$. In contrast, since $[V_i,V_j]=\psi^\dagger[\Gamma_i,\Gamma_j]
\psi$, the matrix element of nested commutators of $V_i$ is of order
$\lambda^mN$. We may therefore identify the common magnitude
$\mu$  with this number $N$, if $\lambda=O(1)$.
Since this nested commutator is still given by a one-body operator, we
may interpret the irreducible 
or quasi-particle amplitude $I_m$ as describing emissions of the
$m$ bosons from the same constituent in the source.

As discussed in the Introduction, we may now conclude that the 
amplitude $A_n$ for the inelastic
reaction
\be
\pi_1+S\to \pi_2+\pi_3+\cdots+\pi_n+S,
\labels{reac}\ee
where the non-relativistic source $S$ is either a heavy nucleus or a baryon
with $N\gg 1$ colors, is of order $\lambda^nN$, instead of
$(\lambda N)^n$ when the 
reaction is described by tree amplitudes like Figs.~1 and 3.

In particular, for $n=1$, we obtain the effective coupling of the pion
to the source to be $g=\lambda N$. If the source $S$ is a baryon with
a large color $N$, it is known \cite{LARGEN} that $\lambda\sim 1/\sqrt{N}$
so the Yukawa coupling constant is $g\sim\sqrt{N}$. The interaction
is then very strong. Under such a circumstance, each Feynman tree diagram
is of order $(\lambda N)^n\sim N^{n/2}\gg 1$, and loop diagrams will
be even larger. There seems to be
 no reason at all to be able to calculate the
process using tree diagrams alone, yet doing so
fairly realistic baryons can be obtained \cite{BARYONS}. What happens
is that although individual tree diagrams are large, their sum gives
rise to a small amplitude $A_n\sim \lambda^nN\sim N^{1-n/2}$,
so loop diagrams computed from $A_n$ are negligible. Moreover,
for large $N$, meson-meson direct couplings are small, so sums of tree
diagrams like Figs.~1 and 3 indeed dominates the whole reaction process in
\eq{reac} \cite{LL2}. 

\section{Parton-Parton Diffractive Scattering}
In order to appreciate what decomposition can do
for this process, let us first summarize
what is known about the QCD calculation of this amplitude, 
at high energy $\sqrt{s}$ and small momentum-transfer
$\Delta=\sqrt{-t}$.

Even assuming the QCD coupling $\alpha_s$ to be very small, 
each loop integration is capable of producing a
factor $\ln s$, so the additional factor for each loop
is likely to be $\alpha_s\ln s$ and not just $\alpha_s$.
If $\alpha_s\ln s=O(1)$ 
which we shall assume from now on, loops of every order must be computed,
and then summed.
Such a difficult task can usually be attempted only
in the leading-log approximation, where the highest power of $\ln s$
is kept at each order. This is equivalent to keeping
only the lowest power of $\alpha_s$ for fixed $\alpha_s\ln s$. 

Such leading-log calculations have been carried out \cite{BFKL}. 
The dominant contribution is of order $\alpha_s$
(for fixed $\alpha_s\ln s=O(1)$), and is mediated by the exchange of 
a color-octet object known as the 
reggeized gluon, or Reggeon for short. For a truly elastic scattering
one needs a color-singlet exchange. That amplitude is of order
$\alpha_s^2$ and the effective object being exchanged
is known as a BFKL Pomeron.
The dependence on $\alpha_s\ln s$ is also known. This dependence
 leads to a total cross section $\sigma_{tot}(s)$
growing like $s^{12(\ln 2)\alpha_s/\pi}$. 
Extrapolated to a large $s$, such a power growth violates unitarity
and the Froissart bound, which forbids any total cross section to grow
faster than $\ln^2s$. 

This problem arises because we extrapolate the BFKL Pomeron amplitude,
correct for $\alpha_s\ll 1$ and $\alpha_s\ln s=O(1)$, 
beyond its region of validity to very large
$s$. To cure the problem
we must add the necessary
subdominant terms before extrapolation.
This is not unlike extrapolating the first
Born approximation valid for weak coupling, to a strong-coupling
regime where contributions
from higher Born terms must be included. 
At high energy this task is somewhat simplified
because there is an impact-parameter
representation for the amplitude
\be
A(s,\Delta)&=&2is\int d^2be^{i{\bf \Delta}\.{\bf b}}\tilde A(s,b),\nn\\
\tilde A(s,b)&=&1-e^{2i\delta(s,b)}.
\labels{exp}\ee
Born approximation corresponds to small
phase shift $\delta(s,b)$, whence the Born amplitude 
$\tilde A(s,b)$ in the impact-parameter
space is proportional to $\delta(s,b)$.
Higher Born approximations are given simply by powers of the
phase shift. To simulate the problem encountered by the 
BFKL Pomeron, onsider an interaction
mediated by the exchange of a spin-$J>1$ particle with an interaction
range $\mu^{-1}$. The phase shift at 
large $b$ has the form
$\delta(s,b)=c(s/s_0)^{J-1}\exp(-\mu b)$. 
It is small if $c\ll 1$ and $s\sim s_0$. 
When $s$ becomes large, the
phase shift is no longer small, so the full expression \eq{exp} has
to be used. The Born approximation violates the Froissart bound 
when extrapolated to large $s$, but
the full expression can be shown not to \cite{DKL}.

This would be a good way to cure the problem of the BFKL Pomeron
if it  could be
interpreted as a phase shift.
For this scenario to be true it is necessary to
demonstrate that the correction terms are given by powers of the
phase shift. Unfortunately this is 
difficult to do, for factorization is not easy to
prove by the usual means, and in any case in the region
$\alpha_s\ll 1$ and $\alpha_s\ln s=O(1)$ where we have control,
the correction terms are small and difficult to calculate.
However, as discussed in the previous sections, multi-quasi-particle
amplitudes do factorize, and within that formalism subdominant terms
can also be calculated just by multiplying a number of dominating
terms. So with the help of the decomposition formula
it is at least hopeful that we may be able to interpret the BFKL-Pomeron
amplitude as a phase shift.

The reality is actually more complicated. We shall summarize here what transpires
and the detail can be found in Ref.~\cite{DKL}.
First we must find a decomposition formula suitabke for the parton-parton
amplitude, by applying the result of Sec.~2 to each of the two energetic
partons. In the region of interest, 
$\alpha_s\ll 1$ and $\alpha_s\ln s=O(1)$, it turns out that
the amplitude $\widetilde A(s,b)$ can be decomposed into sums of products
of irreducible amplitudes $2i\delta_k(s,b)$. This is analogous to the
situation in Sec.~2 where
the tree amplitude $\sum_nA_n$ can be decomposed into sums of products
of the irreducible amplitudes $Q[\s]$. The difference is that
each $Q[\s]$ is a single-quasi-particle amplitude, but
$\delta_k(s,b)$ is characterized by
 $k$ mutually interacting quasi-particles
being exchanged in the $t$-channel,
as illustrated in Fig.~4. The difference arises because gluon interaction
was ignored in the tree amplitude considered in Sec.~2.
When included,
gluons within the same
quasi-particle as well as those in different quasi-particles
may interact, which is why single-quasi-particle amplitudes do not
necessarily factorize anymore. 
Nevertheless, each quasi-particle is still a
color octet, and each would be associated with a common magnitude
of the order of $\mu=\alpha_s$, in the sense that 
\be
\delta_k(s,b)=O(\alpha_s^k). 
\labels{dk}\ee
The factorized multi-quasi-particle amplitudes of all orders can be
summed up to an exponential form
to yield the impact-parameter representation
\eq{exp}, with the phase shift given by
\be
\delta(s,b)=\sum_{k=1}^\infty\delta_k(s,b).
\labels{del}\ee

To connect this with the BFKL-Pomeron amplitude, let us first use \eq{dk}
to conclude 
that the dominant amplitude for $\widetilde A(s,b)$, in the region
$\alpha_s\ll 1$ and $\alpha_s\ln s=O(1)$, comes from $\delta_1$
and is of order $\alpha_s$. In this regime, $\widetilde A(s,b)\simeq
-2i\delta_1(s,b)$ is mediated by the exchange of a single color-octet
quasi-particle. As stated in the beginning of this section, such a
dominant octet amplitude is the Reggeon amplitude, so we can simply
identify the totality of single quasi-particles with a Reggeon.
To obtain a color-singlet amplitude, at least two quasi-particles
must be exchanged, in which case $\widetilde A(s,b)$ is dominated
by a combination of $\delta_1^2$ and $\delta_2$, both of order
$\alpha_s^2$. The first term describes two non-interacting quasi-particles,
and the second two mutually interacting ones. We may therefore
identify the BFKL-Pomeron amplitude with the color-singlet component
of this two-quasi-particle amplitude. Thus Feynman diagrams can be summed
up to give a unitary formula \eq{exp} respecting the
Froissart bound,
and the phase shifts $\delta_1$ and $\delta_2$
can be obtained from the leading-log Reggeion and the
BFKL-Pomeron amplitudes. We might want to drop 
the $\delta_k$
for $k>3$ on the grounds that they are subdominannt
according to \eq{dk}, but they may no longer be small
when $s$ is extrapolated.

\section{Recoil Correction}
So far the propagator has been taken to be
\eq{propx} where all recoils are ignored. 
Motivated by the need of the
Landau-Pomeranchuk-Migdal (LPM) effect 
\cite{LPM}, we shall study in this section how to include recoil.
 As mentioned in the Introduction, the LPM effect
 is of order $1/p^+$, so recoil correction to this order
has to be taken into account. 
In what follows we shall discuss how recoil can be incorporated into the
propagator, how this correction affects abelian factorization, and finally
how the new factorization can be summed up into an exponential form for
the wave function. We will not carry this forward to discuss the LPM
effect as such because discussions can be found elsewhere \cite{LPM,DL}.
The factorization and the subsequent rendering of
an  exponential form is important because phase information, which
is crucial for the radiation suppression contained in the LPM effect,
is contained in the exponent. The treatment below parallels the 
discussion in Ref.~\cite{DL}, but it is presented here in the
configuration space, making it both
algebraically simpler and intuitively more evident.

\subsection{Propagator}
Let us return to the inverse propagator in its exact form,
 $(p+K)^2-m^2+i\e=(p+K)^+K^--k_\perp^2+i\e$. We proceed to change
$(p+K)^+$ to $p^+$, on the grounds that $p^+\gg K^+$, but make no
further approximations at this point. 
In that case instead of \eq{prop} we have $[p^+K^--k_\perp^2+i\e]^{-1}$,
whose Fourier transform gives the propagator in configuration
space to be
\be
\widetilde P(x)&=&{1\over (2\pi)^4}\int d^4(p+K)e^{-i(p+K)\.x}
{1\over (p+K)^2-m^2+i\e}\nn\\
&\simeq& -{i\over (2\pi)^2p^+}\int d^2k_\perp
e^{-ip\.x-ik_\perp^2(x^+/p^+)-ik_\perp\.x_\perp}\d(x^-)\theta(x^+)\nn\\
&=& -{1\over 4\pi x^+}e^{-ip\.x+i(p^+/4x^+)x_\perp^2}\d(x^-)\theta(x^+).
\label{propxx}\ee
The Gaussian function in $x_\perp$ has a rms width proportional to
$x^+/p^+$, so we may think of it as a result of random walk growing with
the lightcone time $x^+$. To the first subleading order, 
we may approximate it by
\be
-{1\over 4\pi x^+}\exp\[i(p^+/4x^+)x_\perp^2\]\simeq
-{i\over p^+}\(\d^2(x_\perp)+{x^+i\over p^+}\nabla^2\delta^2(x_\perp)\),
\nn\\
\labels{deriv}\ee
where $\nabla^2$ is the two-dimensional Laplacian in the transverse
variables.
The first term in \eq{deriv} is the recoilless contribution
given in \eq{propx}. The second term gives
the recoil correction needed for the LPM effect.

\subsection{Leading-Order Wave Function}
Consider a charged particle with relativistic momentum $p$ moving in a
background vector potential $A^\mu(x)$. The $i$th scattering 
vertex is given by $V_i=2p\.A(x_i)=p^+A^-(x_i)$. 
In this subsection we will review  how to compute the wave function of
the charged particle in the leading order. In the
next section we will study how the first subleading corrections
can be included. 

The outgoing
wave function $\psi(x_0)$ can be obtained from the configuration-space
off-shell amplitude of \eq{anx} by
\be
\psi(x_0)=e^{-ip\.x_0}\sum_{n=0}^\infty {1\over n!}\int
\prod_{i=1}^nd^3x_i\widetilde A_n,
\labels{wf}\ee
where $n!$ is the symmetry factor for identical bosons.
Since the vertex is abelian, the result of the 
permutation summation in \eq{anx}
is just to replace the $\theta(x^+)$ factors of \eq{propx}
in all the propagators by $\prod_{i=1}^n\theta(x_i^+-x_0^+)$.
For simplicity we shall drop the subscript 0 in the coordinates
of the wave function, and write $\h A^-(x)$ as $V(x)$.
Then the outgoing wave function in the leading order is
\be
\psi^{(0)}(x)&=&e^{-ip\.x}\sum_{n=0}^\infty {(-i)^n\over n!}
\chi_0(x)^n\nn\\
&=&\exp\(-ip\.x-i\chi_0(x)\),\labels{psi0}\ee
where
\be
\chi_0(x)&=&\h\int_{x^+}^\infty d{x^+}'A^-({x^+}'x^-x_\perp)
\equiv \int_{x^+}^\infty d{x^+}'V(x')
\labels{chi0}\ee
describes the phase shift it accumulates as the particle moves  down
its path. The integration variable $x'$ is understood to have 
the same minus and perpendicular components as $x$. The same convention
will be used for the integration variables $x'$ and $x''$ later in
equation \eq{psi1n3}.

\subsection{Subleading Correction}
When the subleading term of propagator \eq{deriv} is taken
into account, both factorization and exponentiation become more
involved, owing to the presence of the Laplacian operator in 
$\nabla^2\delta^2(x_{i\perp}-x_{i+1\perp})$, as well as the 
additional factor
$(x_i^+-x_{i+1}^+)$ in the correction term. The saving grace
is that we need to compute
the correction only to the first subleading order $1/p^+$, so these 
corrections come in only linearly. 
Then the correction term is
\be
\psi^{(1)}(x)&=&{i\over p^+}e^{-ip\.x}\sum_{n=0}^\infty{(-i)^n
\over n!}\int_{x^+}^\infty \prod_{k=1}^ndx_k^+\ {\cal C}_n,
\labels{psi1}\ee
where
\be
{\cal C}_n&=&\sum_{i=1}^n\biggl[
\nabla^2\(\prod_{k=1}^{i-1}\theta(x_k^+-x_i^+)V(x_k^+)\)\nn\\
&&\theta(x_i^+-x_{i+1}^+)(x_i^+-x_{i+1}^+)\prod_{l=i}^nV(x_l^+)\biggr]
\nn\\
&=&\sum_{i=1}^n(x_i^+-x^+)\nabla^2V(x_i^+)\prod_{j\not=i}V(x_j^+)+\nn\\
&&2\sum_{i<j}\nabla V(x_i^+)\.\nabla V(x_j^+)\theta(x_i^+-x_j^+)(x_j^+-x^+)\prod_{l\not=i,j}V(x_l^+).\nn\\
&&\labels{cn}\ee
Thus
\be
\int_{x^+}^\infty \prod_{k=1}^ndx_k^+\ {\cal C}_n&=&
\chi_0^{n-2}(x)\[n\chi_0(x)\chi_2(x)+n(n-1)\chi_1(x)\],\nn\\
\chi_2(x)&\equiv&\int_{x^+}^\infty d{x^+}'({x^+}'-x^+)\nabla^2V({x^+}')
\nn\\
&=&\nabla^2\int_{x^+}^\infty d{x^+}'\chi_0(x'),\nn\\
\chi_1(x)&\equiv&\int_{x^+}^\infty dx^{+'}({x^+}'-x^+)\nabla
V(x')\.
\int_{x_{+}'}^\infty d{x^+}''\nabla V(x'')\nn\\
&=&\int_{x^+}^\infty d{x^+}'\[\nabla\chi_0(x')\]^2.
\labels{psi1n3}\ee
Summing up $n$,
we find to the accuracy of  order $1/p^+$ that the outgoing wave
function is given by
\be
\psi(x)&=&\psi^{(0)}(x)+\psi^{(1)}(x)\nn\\
&=&\exp\[-ip\.x-i\chi_0(x)
-{i\over p^+}\(\chi_1(x)+i\chi_2(x)\)\].
\labels{psif}\ee 

\acknowledgements
This research is supported in part by the Natural Sciences and
Engineering Research Council of Canada and the Fonds pour la
formation de Chercheurs et l'Aide \`a la Recherche of Qu\'ebec.

\begin{figure}[h]
\vspace*{4cm}
\includegraphics{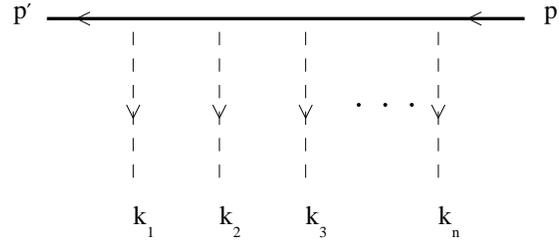}
\vspace*{3cm}
\caption[]{A tree diagram showing bosons of momenta $k_i$
being emitted from an energetic source with initial momentum
$p$ and final momentum $p'$.}
\end{figure}

\vspace{5cm}

\begin{figure}[h]
\vspace*{2cm}
\includegraphics{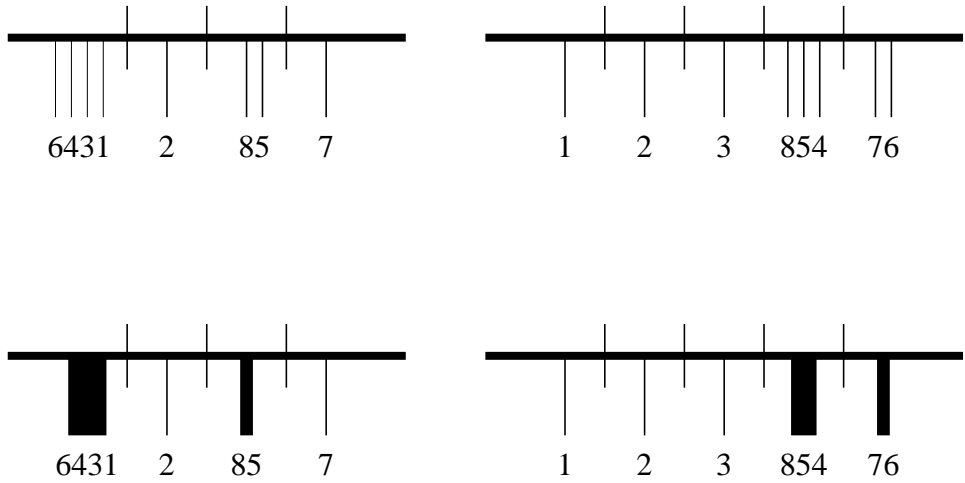}
\vspace*{3cm}
\caption[]{Two $n=8$ non-abelian cut diagrams showing where
cuts should appear. For each of them, a simplified version
is given in which every quasi-particle is shown as a thick
vertical line.}
\end{figure}

\newpage

\begin{figure}[h]
\vspace*{5cm}
\includegraphics{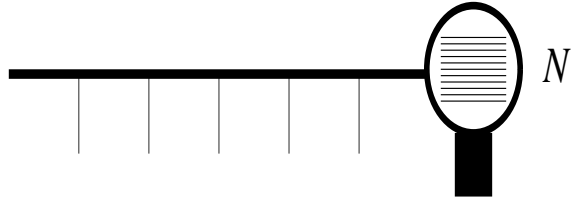}
\vspace*{2cm}
\caption[]{A tree diagram whose source has $N$ constituents.}
\end{figure}

\vspace{5cm}

\begin{figure}[h]
\vspace*{2cm}
\includegraphics{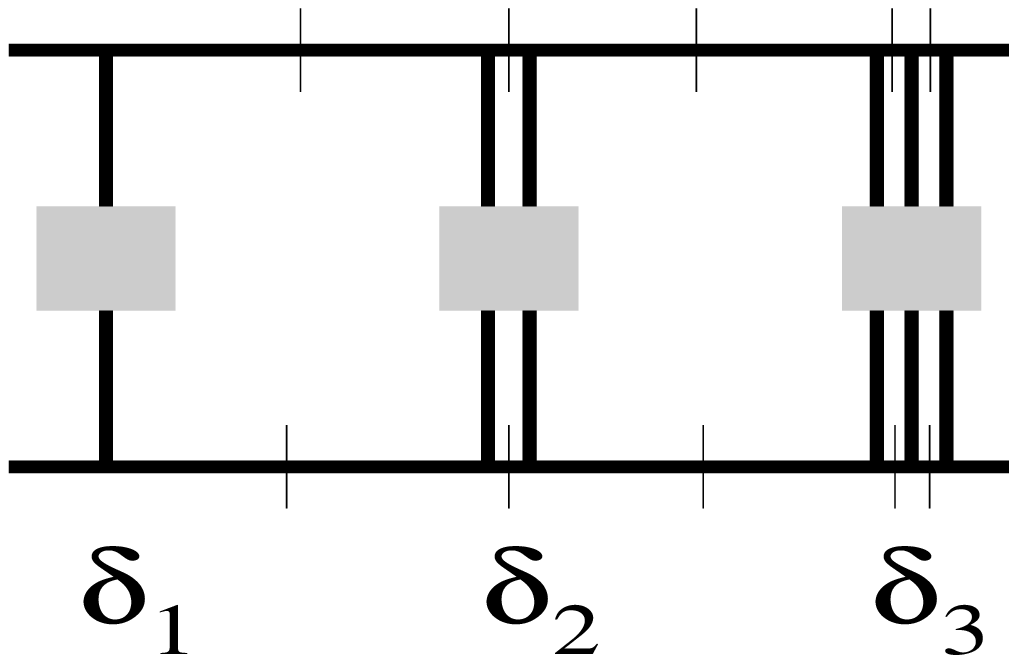}
\vspace*{4cm}
\caption[]{The decomposition of a parton-parton scattering amplitude
into products of irreducible amplitudes $\delta_k$, where $k$ indicated
the number of mutually interacting quasi-particles (thick vertical line)
being exchanged.}
\end{figure}

\vspace{3cm}


\begin{thebibliography}{9}
\bibitem{LL1LJMP}C.S. Lam and K.F. Liu, {\it Nucl. Phys.} 
{\bf B483} (1997) 514;
C.S. Lam, {\it J. Math. Phys.} {\bf 39} (1998) 5543.
\bibitem{EIK}R. Torgerson, {\it Phys. Rev.} {\bf 143} (1966) 1194;
H. Cheng and T.T. Wu, {\it Phys. Rev.} {\bf 182} (1969)
1868, 1899; M. Levy and J. Sucher, {\it Phys. Rev.} {\bf 186} (1969) 1656.
\bibitem{STERMAN} G. Sterman, {\it An Introduction to Quantum
Field Theory'}, (Cambridge University Press, 1993).
\bibitem{PESKIN} M.E. Peskin and D.V. Schroeder, {\it `An Introduction
to Quantum Field Theory'}, (Addison-Wesley, 1995).
\bibitem{WEINBERG}S. Weinberg, {\it `Quantum Theory of Fields'}
(Cambridge University Press, 1995).
\bibitem{GLAUBER}R.J. Glauber, in {\it `Lectures in Theoretical Physics'},
ed. W.E. Brittin and L.G. Dunham (Interscience, New York, 1959, vol. 1).
\bibitem{CW}H. Cheng and T.T. Wu, {\it
`Expanding Protons: Scattering at High Energies'}, (M.I.T. Press, 1987).
\bibitem{MV} L. McLerran and R. Venugopalan, {\it Phys. Rev. D} {\bf 49} (1994)
2233; {\bf 49} (1994) 3352; {\bf 50} (1994) 2225. 
\bibitem{LM} C.S. Lam and G. Mahlon, hep-ph/9907281.
\bibitem{LARGEN}G. 't Hooft, {\it Nucl. Phys.}
{\bf B72} (1974) 461;
E. Witten, {\it Nucl. Phys.} {\bf B160} (1979) 57;
S. Coleman, Erice Lectures (1979).
\bibitem{LL2}C.S. Lam and K.F. Liu, {\it Phys. Rev. Lett.}
{\bf 79} (1997) 597.
\bibitem{BFKL}For a review, see for example 
L.V. Gribov, E.M. Levin, and M.G. Ryskin,
{\it Phys. Reports} {\bf 100} (1983) 1;
V. Del Duca, hep-ph/9503226;
L.N. Lipatov, {\it Phys. Reports} {\bf 286} (1997) 131;
E. Levin, hep-ph/9808486.
\bibitem{DKL}Y.J. Feng and C.S. Lam, {\it Phys. Rev. D}
{\bf 55} (1997) 4016; R. Dib, J. Khoury, and C.S. Lam, 
hep-ph/9902429
(to appear in {\it Phys. Rev. D}).
\bibitem{LPM}L.D. Landau and I.J. Pomeranchuk, {\it
Dokl. Akad. Nauk. SSSR} {\bf 92} (1953) 535, {\bf 92} (1953) 735;
A.B. Migdal, {\it Phys. Rev.} {\bf 103} (1956) 1811;
R. Blankenbecler and S.D. Drell, {\it Phys. Rev. D}
{\bf 53} (1996) 6265; R. Baier, Yu.L. Dokshitzer,
A.H. Mueller, S. Peign\'e, and D. Schiff, 
{\it Nucl. Phys.} {\bf B483} (1997) 291;
 P.L. Anthony et al., {\it Phys. Rev Lett.}
{\bf 75} (1995) 1949; 
V.N. Baier and V.M. Katkov, {\it Phys. Rev. D} {\bf 57} (1998) 3146;
U.A. Wiedemann and M. Gyulassy, hep-ph/9906257.
\bibitem{FHL}Y.J. Feng, O. Hamidi-Ravari, and C.S. Lam,
{\it Phys. Rev. D} {\bf 54} (1996) 3114.
\bibitem{BARYONS}R.F. Dashen, E. Jenkins and A.V. Manohar, {\it Phys. Rev. D} {\bf 49} (1994) 4713; M.A. Luty and J. March-Russell, {\it Nucl. Phys.} {\bf B426} (1994); R.F. Dashen, E. Jenkins and A.V. Manohar, {\it Phys. Rev. D} {\bf 51} (1995) 3697.
\bibitem{DL}I.M. Dremin and C.S. Lam, {\it Mod. Phys. Lett.} {\bf A13}
(1998) 2789.

\end{thebibliography}
\end{document}